\documentstyle[11pt,aaspp4]{article}

\begin{document}
\title{Mid-Infrared Emission from E+A Galaxies in the Coma Cluster}

\author{
A.~C.\ Quillen\altaffilmark{1}$^{,2}$,
G.~H.~Rieke$^1$,
M.~J.~Rieke$^1$,
N.~Caldwell\altaffilmark{3}
\& C.~W.~Engelbracht$^1$
}
\altaffiltext{1}{Steward Observatory, The University of Arizona, Tucson,
AZ 85721}
\altaffiltext{2}{aquillen@as.arizona.edu}
\altaffiltext{3}{F.~L.~Whipple Observatory Smithsonian Institution Box 97,
Amado, AZ 85645; caldwell@flwo99.sao.arizona.edu}

\begin{abstract}

We have used ISO to observe at 12$\mu$m seven E+A galaxies 
plus an additional emission line galaxy, all in the Coma cluster. 
E+A galaxies lacking narrow emission lines
have 2.2$\mu$m to 12$\mu$m flux density ratios or limits similar to
old stellar populations (typical of early-type galaxies).
Only galaxies with emission lines have enhanced 12$\mu$m
flux density.  Excess 12$\mu$m emission is therefore
correlated with the presence of on-going star formation or
an active galactic nucleus (AGN).

From the mid and far infrared colors of the brightest galaxy in our sample,
which was detected at longer wavelengths with IRAS, we
estimate the far infrared luminosity of these galaxies.
By comparing the current star formation rates with previous rates
estimated from the Balmer absorption features, we divide the galaxies
into two groups: those for which star formation has declined significantly
following a dramatic peak $\sim$ 1 Gyr ago; and those with a significant
level of ongoing star formation or/and an AGN. There is no strong 
difference in the spatial distribution on the sky between these two groups. 
However, the first group has systemic velocities above the mean cluster 
value and the second group below that value.  This suggests that 
the two groups differ kinematically.

Based on surveys of the Coma cluster in the radio, the IRAS sources, and galaxies detected in 
H$\alpha$ emission, we sum the 
far infrared luminosity function of galaxies in the cluster.
We find that star formation in late type galaxies is 
probably the dominant component of the Coma cluster far infrared luminosity. 
The presence of significant emission from intracluster dust is not yet firmly established. 
The member galaxies also account for most of the far infrared output 
from nearby rich clusters in general. We update
estimates of the far infrared luminosities of nearby, rich clusters and
show that such clusters are likely to undergo luminosity evolution
from $z = 0.4$ at a rate similar to, or faster than, field galaxies.

\end{abstract}

\keywords{
galaxies: clustering
-- evolution: galaxies
-- infrared: general
-- clusters: individual: Coma
-- clusters: general
}

\section {Introduction}

A subset ($\sim 30\%$) of the members of moderate redshift galaxy
clusters have abnormally blue colors (e.g.~\cite{but78};
\cite{but84}; \cite{cou87}; \cite{dre82}), 
coined the ``Butcher-Oemler effect''.  Spectroscopy
(e.g.~\cite{dre83}) of these galaxies confirms cluster membership
and shows spectral features such as Balmer absorption lines,
indicating the presence of a post starburst population,
or/and strong narrow emission lines,
indicating ongoing active star formation or an AGN.
Elliptical or S0 galaxies with strong Balmer absorption features in their spectra are commonly denoted E+A galaxies.
Both for the Butcher-Oemler clusters (\cite{cou98}) and more distant ones
(\cite{dre94}; \cite{rakos}) 
many of the blue galaxies are spirals, a population which is relatively
rare in nearby rich clusters. 
The discovery (\cite{vigroux}, \cite{cald93}) 
of E+A and non-spiral emission line 
galaxies in the Coma cluster, 
but at a redshift of only $z=0.023$ rather than the $0.2-0.5$ 
of the Butcher-Oemler clusters,
makes it possible to study in greater detail the processes resulting in
star formation in clusters.

The mid and far infrared luminosity
is a particularly useful measure in the study
of galaxy evolution because it is a diagnostic of recent
or ongoing star formation.
The ISO satellite (\cite{kes96})\footnote{Infrared Space Observatory (ISO) is an ESA project funded by ESA Member States (especially the PI countries: France, Germany, The Netherlands, and the United Kingdom) with the participation of ISAS and NASA} has allowed us to observe the mid-infrared outputs of
galaxies in nearby clusters.
Of the 182 early-type galaxies surveyed with spectroscopy from the Coma cluster
(by \cite{cald93} and \cite{cald97}),
20 showed Balmer absorption features or narrow emission lines.
In this paper we present ISO observations at $12 \mu$m of seven
of these 20 galaxies and include one control or `normal'
cluster galaxy lacking these spectral features.
One additional late-type galaxy (D100) appears in the same
field as one of the post starburst or E+A galaxies (D99).

Ultimately the rate of star formation in any environment
must be related to the availability of cold gas from
which stars can form. The hierarchical picture of cluster 
formation holds that this reservoir of gas
is replenished by spiral galaxies coming into the cluster.
Merging of subclusters or infalling galaxy groups would then 
result in a period of elevated star formation in the cluster galaxies.
Star formation in individual galaxies
is triggered by the cluster
tidal field, perturbations from nearby galaxies (also referred
to as galaxy harrassment; \cite{moore96}), or interactions with the intracluster
medium.  Incoming galaxies are expected to
have an initial burst of star formation, after which
galaxy harrassment and the cluster tidal
field result in stripping, heating of the stellar velocity dispersion, and 
possibly a lower level of continuing star formation over a longer
timescale (e.g.~\cite{moore}). 

These processes are expected to take place in high redshift
clusters. However, even if major mergers of gas rich subclusters are common
in the higher redshift clusters, 
it would be difficult to separate individual cluster infall events. 
Evidence for such a merger is present in the Coma cluster: 
it has two major ellipticals and two peaks in the X-ray flux density
and radio source counts (\cite{kim}), 
one of which is associated with a number of galaxies with E+A and 
emission line spectral characteristics (\cite{cald93}). 
Kinematic differences observed in the galaxy populations can be
interpreted in terms of interactions between the subcluster
and the main body of the cluster and the later
infall of field galaxies (\cite{colless}).
The association of E+A galaxies with cluster sub-structure
would be consistent with an initial star formation burst period beginning
about a Gyr ago when the subclusters first began to interact. 
The possibility of a merger can be probed further by examining 
the star formation histories of these individual ``abnormal'' Coma galaxies. 

By comparing current star formation rates estimated with our ISO observations
with those $\sim 1$ Gyr ago estimated
from the Balmer absorption lines we can study the recent
star formation histories in these galaxies.
Our sample consists of a representative group of Coma cluster galaxies with 
``abnormal'' spectral features such as Balmer absorption lines or/and emission 
lines. From their star formation histories
we can estimate timescales for star formation in the cluster
associated with a triggering event such as the infall of a sub-cluster.

IRAS and ISO studies of clusters have resulted in detection
of extended far infrared emission 
(\cite{hu_}, \cite{stickel_}, \cite{wise_}).  
Part of this emission has been attributed to intracluster dust, 
as modeled by \cite{dwek} to be consistent with 
reports of possible extinction through the Coma cluster. 
The spatial resolutions of IRAS and ISO were not sufficient to
determine whether the cluster far infrared emission is smoothly
distributed, as expected from intracluster dust, or associated
with individual galaxies. Recent observational surveys of the Coma cluster
in the radio (\cite{kim}) and H$\alpha$ emission (\cite{gavazzi_})
and the compiled IRAS identifications of  \cite{wang}
make it possible to determine the infrared luminosity function and 
estimate the cluster luminosity.  Here we investigate
the possibility that galaxies in the cluster
are the dominant source of infrared emission and 
could result in the color signature that has been attributed 
to intracluster dust (e.g.~\cite{stickel_}). 

This paper is organized as follows. In \S 2 we present the 12$\mu$m 
ISO observations of the galaxies. To determine whether the 12$\mu$m flux density is above
that expected from a quiescent old stellar population
we also observed these galaxies at K band (2.2$\mu$m).
The 12 and 2.2$\mu$m flux densities are compared in \S 3.1 and the set of galaxies
with enhanced 12$\mu$m emission is identified.
In \S 3.2 we check our estimates for the far infrared luminosities
by comparing the 12$\mu$m flux densities of our brightest galaxies
to IRAS or radio flux densities. 

In \S 4.1 star formation histories for these galaxies are discussed.
We compare the current star formation rate
estimated from the infrared emission to that
derived from H$\alpha$ emission.  We then
compare the current star formation rates to the
past rates inferred from the Balmer absorption lines.
We identify two groups of galaxies in our sample:
those having current star formation rates that
are substantially lower than that required to form their $\sim 1$ Gyr old
populations and those with significant ongoing star formation
or/and AGN activity.  We then compare the spatial and velocity distribution
of the two groups. In \S 4.2 by considering number counts
of radio sources, IRAS identifications and H$\alpha$ detections
we constrain the form of the far infrared luminosity
function of galaxies in the cluster.  We then estimate the total
far infrared luminosity of the Coma cluster.
In \S 4.3 we discuss the complications in measuring emission by intracluster dust because of the large contribution by cluster galaxies. In \S 4.4 we
re-examine the luminosity evolution of cluster galaxies using
measurements of distant clusters from \cite{kelly}.

\section {Observations}

Throughout this paper, we refer to galaxies by NGC number or,
when they have none, by their \cite{dre80} catalog or D number.
We adopt a Hubble constant of $H_0 = 75$ km s$^{-1}$ Mpc$^{-1}$.

\subsection{ISOCAM imaging at 12$\mu$m}

Images of 9 fields each centered on a Coma cluster galaxy
were taken with ISOCAM (\cite{ces96})
in the imaging mode with $6.0''$ per pixel
and using the LW10 filter (8.6-14.4$\mu$m).
Each field was observed for a
total of 115 exposures each 5.04s long,
taken in a $3\times 3$ raster on
the sky, offsetting by $36''$ between points.
Each point was observed for a total of $\sim 50$ seconds
with additional time for stabilization on the first
position.  The total integration time per galaxy was 9.7 minutes.

In each set of observations, we discarded the first 15 of
115 exposures because the array had not yet stabilized.  Based
on the subsequent flat level of the sky,
we judged the array to be stabilized after this time.
Cosmic ray glitches were removed using the `MM'  multi-resolution
spatial and temporal routine in the CAM Interactive Analysis (CIA)
package.\footnote{CAM Interactive Analysis is a joint development by the ESA
astrophysics division and the ISOCAM consortium}
Because of the extended total integration time, low frequency
noise was observed in each pixel.
To suppress this noise,
we removed slowly varying
third order polynomials in time from each pixel signal,
excluding locations of bright source emission.
Following this procedure, flat fields were constructed
from the images themselves using an interpolator program
to fit a smooth function of time to the entire data cube.
The final mosaic was then constructed from the nine mosaic positions
with a least-squares fit to each pixel in the images, after shifting them
according to position on the sky.  The final images are shown in Figure 1.

Images were flux calibrated using sensitivity parameters listed
in later versions of the ISOCAM data pipeline that are probably
accurate to within $\pm 20\%$ (\cite{ces96}).  Since the images used
were observed after the array stabilized, the calibration should
be accurate to this level.
The resulting images have a pixel to pixel standard deviation
of $\sim 0.04$ mJy per $6''$ pixel in the central $25 \times 25$ pixels.
By removing some of the low frequency noise
from the images we decreased the pixel to pixel standard deviation
by a factor of $\sim 5$ from the Automatic Analysis product.

\subsection{Near-infrared J and K imaging}

Near-infrared J and K$_s$ broad band images were
obtained on May 8, 1998 at the 90$''$ (2.3m) Bok Telescope on
Kitt Peak with a
$256\times 256$ NICMOS3 near-infrared camera.
On source total exposure times in each filter were 8 minutes for
NGC 4865, 4 minutes for NGC 4853 and 9 minutes for the remaining
galaxies. Observing conditions were photometric so images were calibrated
by comparing to standards from \cite{elias}.
Aperture photometry is listed in Table 1.
$9.6''$ diameter apertures were used to measure the J-K color.
K band magnitudes were measured in a larger $36''$ aperture, with nearby
stars and galaxies removed, to compare to the flux densities observed at 12$\mu$m.
The near-IR J-K colors are typical of normal galaxies, except
in the case of D45 which is significantly bluer
than the other galaxies, indicating the presence
of younger stellar population or a lower metallicity.

\section{Results}

\subsection{Galaxies detected at 12 microns}

In the 12$\mu$m images (displayed in Fig.~1), 6 of 9 galaxies
are clearly detected at a level of greater than a mJy. We also 
observed the control (or normal) galaxy,
NGC~4865, as well as a bright E+A galaxy
which had been detected at longer wavelengths
with IRAS, NGC~4853. Detected galaxies were seen at different 
locations in the data cube so we are confident that we are not misinterpreting cosmic ray hits.
Because of the remarkable pointing performance of ISO
(absolute pointing error less than a few arcsecs, \cite{kes96})
we are confident that the images are centered
at the positions of the galaxies.
All detected galaxies except for D99 and D100 were located
at the center of the image.
The galaxies D99 and D100 are $\sim 20''$ apart and so
can be viewed in the same image, which is centered on
D99.  However the bright source detected in the 12$\mu$m  image is
to the north-west of the center of the image
and so at the expected location of D100 not D99.
This source is therefore probably
D100, which has narrow emission lines indicating
active ongoing star-formation as well
as Balmer absorption features, rather than
the more quiescent post starburst D99,
which has only Balmer absorption features (\cite{cald96}).
Fluxes at 12$\mu$m measured from these images are tabulated in Table 1.

The brighter sources at 12$\mu$m
(D16, D100, NGC~4865 and NGC~4853) are clearly extended
even at the resolution of our $6''$ pixels.
To illustrate, we have included a point source
in Fig.~1 for comparison to the galaxies.
The fainter galaxies (in B band; all but NGC~4865
and NGC~4853) have scale lengths (measured in B band images)
of $\sim 1.5$ kpc or a few arcseconds
whereas NGC~4853 has a larger scale length of
$\sim 10''$ (\cite{cald96}).  Thus, the galaxies have 
similar extent in the mid infrared and blue. 

\subsection{Comparison of galaxies with and without narrow emission lines}

D16, D44, D45, D94, D99, D100 and D112
all have similar luminosities in B band (see Table 1).
Of this group, the galaxies with narrow emission lines (D16, D44, D45 and D100)
as well as Balmer absorption features were clearly detected at 12$\mu$m. 
The brighter galaxy, NGC~4853, 
also has faint narrow emission lines as well as Balmer
absorption features (\cite{cald96}) and is detected at 12$\mu$m.  
Of the galaxies lacking emission lines (the purely post starburst E+A galaxies
D94, D99, and D112), none were clearly detected.
Bright pixels at the expected location of
these latter galaxies are observed in all three images;
however the detection is statistically significant
only for D112. A bright pixel coincident with this galaxy
was observed in more than one place in the data cube and
the net signal in the final image is 3 times above the standard deviation of
the pixel to pixel noise fluctuations in the central $10\times 10$ pixels.
We tentatively say we have detected D112 at the level of 0.4 mJy
and we have not detected D99 and D94 with upper limits on the flux densities
of 0.3 mJy.

To determine if the galaxies
have excess 12$\mu$m emission, we compare the $2.2\mu$m to $12\mu$m
flux density ratio with that expected from an old
quiescent stellar population. NGC~4865 is a control galaxy for our sample
and has no Balmer absorption  features (\cite{cald93});
it has a flux density ratio $F_{2.2\mu{\rm m}}/F_{12\mu {\rm m}} \sim 6$, 
which is typical of early-type galaxies (\cite{imp86}). 
NGC~4853, D16, D44, D45 and D100 are significantly brighter at 12$\mu$m 
(compared to their 2.2$\mu$m flux densities) than such galaxies. 
Much of the UV emission exciting the low ionization narrow emission 
lines in these galaxies appears to be reradiated by dust at longer
(infrared) wavelengths. 

D16 has high excitation narrow emission lines, no Balmer
absorption features and a broad H$\alpha$ component,
and so is likely to be a Seyfert 1 galaxy
(\cite{cald93}; \cite{cald96}).
Its luminosity at 12$\mu$m would probably
support this interpretation; however, the
12$\mu$m emission is extended so it is likely that
there is some active star formation occurring in this galaxy as well.
D44 is intermediate in emission line characteristics, between
a LINER and a Seyfert 2 (\cite{cald96}),  and is not
bright enough that we can determine whether it is extended
at 12$\mu$m.

The pure post starburst
galaxies D94, D99 and D112 (lacking narrow emission lines)
have 2.2 to 12$\mu$m
flux density ratios (or limits in the case of D94 and D99)
consistent with expectations for old stellar populations.
In such galaxies, the 12$\mu$m emission scales with
the $2.2\mu$m emission and so likely originates from
thick dusty interstellar envelopes that form around
red giant stars due to mass loss (\cite{soi86}; \cite{kna92};
\cite{imp86}). Thus, the pure E+A phase corresponds to a later 
(or post) starburst phase where the brightest, 
most massive stars have already left the main sequence and correspondingly 
the reradiated infrared flux is low. 

\subsection{Comparison with IRAS and radio flux densities for NGC~4853 and D100}


The Coma cluster E+A galaxy NGC~4853 was detected with IRAS with
$F_{60\mu {\rm m}} = 0.65$ Jy, and
$F_{100\mu {\rm m}} = 1.48$ Jy, giving $F_{60\mu {\rm m}}/ F_{100\mu {\rm m}}
\sim  0.44$.  It was not detected at shorter
wavelengths, but from our observed 12$\mu$m flux density, we find that
$F_{12\mu {\rm m}}/ F_{60\mu {\rm m}} = 0.061 \pm 0.01$.
This comparison with IRAS
is valid, since the galaxy scale length is $\sim 10''$ (\cite{cald96}) and
we expect little extended emission past what we see in
our 12$\mu$m image.  The ratio
$F_{12\mu{\rm m}} / F_{60\mu{\rm m}} = 0.06$
is also consistent with the mean value of the IRAS bright galaxy 
sample (\cite{soi89}) and the 12$\mu$m normal galaxy sample of \cite{spinoglio}.
Thus, both the mid and far infrared
colors of NGC~4853 are typical of a normal spiral galaxy.

NGC~4853 and D100 are included in the list of optical identifications
of radio sources in the Coma cluster (of \cite{kim}). 
At 1.4GHz, the flux densities are $2.9\pm 0.3$mJy and $1.6\pm 0.4$mJy, 
respectively (\cite{kim}).
The ratio of far-infrared to radio flux densities for NGC~4853
is comparable (but at the high end of the distribution) of that seen from the
infrared to radio correlation (for the IRAS galaxies of \cite{hel85_}),
and our predicted ratio for D100 coincides with the mean of the distribution.
Thus, D100 and NGC~4853 have
infrared to radio flux density ratios that are typical
of normal galaxies and not particularly low as found for some
spiral galaxies in rich clusters (\cite{anderson}).

\subsection{Far-infrared luminosities}

We estimate the far infrared luminosities for the Coma cluster galaxies as
$\nu F_\nu$ at $60\mu$m
assuming a flux density ratio of $F_{12\mu{\rm m}} / F_{60\mu{\rm m}} = 0.06$
(consistent with observations of NGC~4853).
The resulting luminosity, $L_{IR}$, is approximately equal to the parameter
$L_{FIR}$ of \cite{hel88} based on the 60 and 100$\mu$m flux densities for
a galaxy with a typical ratio $F_{60\mu{\rm m}}/F_{100\mu{\rm m}} = 0.5$.
Infrared luminosities (Table 1) have 
only been estimated for galaxies with enhanced 12$\mu$m emission compared to their 
$2.2\mu$m flux densities.
Most of the galaxies have low infrared to optical flux density ratios
and moderate far infrared luminosities.
The infrared to optical flux density ratios are similar to those
of optically selected galaxies,
which have log$(L_{IR}/L_{opt}) \sim -0.3$
for Shapely-Ames galaxies (\cite{dejong}, where $L_{opt}$
is computed as $\nu F_\nu$ at B band).
There is no obvious correlation between this ratio and the
infrared flux density (as observed in the IRAS bright galaxy sample; \cite{soi89}).  
The lack of correlation may result from our small sample and
the decaying star formation rates in most of these galaxies (\cite{cald96}).

\section{Discussion}

\subsection{Star formation histories}

Here we compare star formation rates estimated from elevated 12$\mu$m
emission to those estimated from H$\alpha$ emission and a previous
epoch inferred from the presence of Balmer absorption lines.
We first consider the more quiescent galaxies. 
From optical absorption lines, \cite{cald96_} estimated that D94, D99 and D112 have $\sim 60\%$ of their light
at $4000\AA$ from a starburst population $0.8-1.3$ Gyr old.
Assuming a Saltpeter IMF, a few times $10^9 M_\odot$ stars
were formed during this episode of star formation (e.g., using
mass-to-light ratios given in \cite{worthey}). These values correspond very 
approximately to star formation at a rate of $\sim$ 5 M$_\odot$/yr
averaged over a Gyr. The absence 
of observed line emission restricts the current
star formation rates to less than $\sim 0.1 M_\odot$/yr
(using a conversion from \cite{kenn83}). 
However, the lack of 12$\mu$m emission, 
corresponding to $L_{IR} \lesssim 10^8 L_\odot$
(using conversion factors discussed in \cite{leth95} and \cite{dev90}), 
limits the current star formation rate even more, $\lesssim 0.03 M_\odot$/yr. 
NGC~4853 has a star formation rate (estimated from
the H$\alpha$ flux) of $\sim 1 M_\odot$/yr (\cite{cald96}),
which is consistent with the far-infrared luminosity of
$7 \times 10^9 L_\odot$ estimated from the $12\mu$m emission.
\cite{cald96_} and \cite{sparke}
estimated an age of $\sim 0.9$ Gyr for a population
contributing $\sim 60\%$ of the light at $4000\AA$, which implies
that $\sim 10^{10} M_\odot$ stars were formed during this burst, 
i.e., a rate of $\sim 10 M_\odot$/yr or higher averaged over a Gyr. 
Thus, the current star formation rates in NGC~4853, D94, D99 and D112
are well below those required to form the $\sim 1$ Gyr old burst populations
responsible for their Balmer absorption features.

D44, on the other hand, has an E+A spectrum as well as
emission lines characteristic of an AGN.
The galaxy is too faint at 12$\mu$m to determine if the emission is resolved.
D44 has absorption lines similar to those of NGC~4853 suggesting
that a population of a few times $10^9 M_\odot$ was formed about a Gyr ago.
However its H$\alpha$ emission (which is primarily due to an AGN)
restricts its star formation rate to less than $0.2 M_\odot$/yr.
The 12$\mu$m emission level restricts its current star formation 
to less than $0.1 M_\odot$/yr. This suggests that this galaxy is less active now than during
the time when it formed its $\sim 1$ Gyr old burst population
responsible for its Balmer absorption features.

D100 has an emission line spectrum similar to HII regions, 
suggesting active ongoing star formation. The emission line fluxes 
correspond to a star formation
rate of $\sim 1 M_\odot$/yr. The optical absorption lines
indicate a $\sim$ $0.5$ Gyr old population, contributing 
$\sim 50\%$ of the light at $4000\AA$ (\cite{cald96}) similar to
the other galaxies and containing a few times $10^9 M_\odot$ stars.
The far infrared flux density predicted from the 12$\mu$m emission
(as described above)
is a factor of a few lower than predicted from the H$\alpha$
luminosity but within the scatter observed by \cite{dev90_}
for the correlation of H$\alpha$ to infrared luminosities.
The higher current star-formation rate and younger post-star burst
population in D100 suggest the current star formation rate is similar to 
that required to form its $0.5$ Gyr old burst population.

D45 has optical absorption spectral features
similar to those of D94, D99, and D112 (\cite{cald93}) and
so probably has a similar young population of a few times $10^9 M_\odot$
and of age $0.5-1.3$ Gyr old.  However this galaxy is irregular,
very blue in optical and near-infrared colors (B-K $\sim 2$) and has
H$\alpha$ emission flux consistent
with a star formation rate of $\sim 0.2 M_\odot$/yr (\cite{cald96})
which roughly corresponds to the far infrared flux density predicted from its
12$\mu$m flux density. As in the case of D100, the star formation rate of D45 is
not necessarily significantly below that required to form
its 1 Gyr old burst population.

We observe that the 12$\mu$m emission in D16, which
contains an AGN,  is extended
and so the galaxy is probably also forming stars.
At least $\sim 50\%$ of the 12$\mu$m flux density is resolved in the image.
Assuming that $\sim 50\%$ of the 12$\mu$m flux density is due to star formation regions
we estimate a star formation rate of $0.1 M_\odot$/yr.
Since D16 does not display Balmer absorption features its activity
is probably recent.

Based on these individual star formation histories 
we can divide our sample galaxies
into two groups: 1) those that have current star formation rates that
are substantially lower than that required to form their $\sim 1$ Gyr old
populations  (NGC~4853, D94, D99, D112, and D44) and
2) those with significant ongoing star formation
(D100, D45) or/and AGN activity (D16) compared to that
occurring $\sim 1$ Gyr ago.
We note no difference in the distribution on
the sky between these two groups. In fact, D99 and D100 are
within the same ISOCAM imaging field and near the cluster
center.   However we do notice a difference
these two groups when we consider their systemic velocities.
The galaxies with ongoing activity have systemic velocities
below the cluster mean velocity, $V=6989$ km/s (\cite{cald93}),
and the galaxies from the more quiescent first group have velocities above
this mean (see Table 1).  \cite{cald93_} noted that the galaxies
with strong Balmer absorption lines have a mean velocity
about 200 km/s above the cluster mean velocity.
Late type galaxies, on the other hand, have a much higher velocity
dispersion than the early type galaxies (\cite{colless}).  
Our first group of galaxies is kinematically consistent with the
the E+A galaxies of \cite{cald93_}, whereas our   
second group of
galaxies would be consistent with the kinematic properties of
the late-type galaxies.
%
%
This would be consistent with the interpretation of \cite{colless_} that 
the E+A galaxies (lacking significant ongoing activity), in the first group, 
are part of a complex which entered the cluster environment $\sim 1$ Gyr ago. 
Galaxies in the second group, with ongoing and more recent activity,
are likely to have entered the cluster
as part of different complexes and at later times.
The kinematic differences between the two groups suggest that 
star formation triggered by the cluster decays on a timescale of order 
a Gyr (consistent with simulations of \cite{moore}).

\subsection{Contributions to the total far infrared flux of the Coma cluster}

Recent surveys conducted in the region of the Coma cluster make
it possible to place constraints on the far infrared luminosity 
function of galaxies in the cluster and so better compute the integrated far infrared
luminosity of the Coma cluster.
We consider three magnitude limited samples:
IRAS, H$\alpha$ and radio.

The total far infrared luminosity of 
of a nearby rich cluster such as the Coma cluster can be
estimated from the total number of galaxies that
are detected at a particular limiting luminosity
and integrating the far-infrared luminosity function
(e.g.,~\cite{kelly_}). To apply this method, we need to constrain 
the slope of the faint end of the galaxy luminosity function.
\cite{kelly} adopted a luminosity function in the
form of two power laws and a slope at the low luminosity end of $-0.8$.
They derived a total integrated flux density at $z = 0.05$ of 2250 mJy at
$60\mu$m for a typical rich cluster.   However, more recent studies
have suggested that the field galaxy luminosity function has a low end
slope near $-0.1$ (\cite{saunders}).
The optical luminosity function in rich clusters has a slope of
about $-0.25$ at its low end (e.g., \cite{dre84}).
It therefore appears likely that the value adopted by Kelly and Rieke
is too steep. Slopes of either $-0.1$ or $-0.25$ would yield an average
$60\mu$m flux density for a rich cluster
at $z = 0.05$ of $\sim 800$ mJy, calculated by correcting
the entry of 2250 mJy in Table 2 of \cite{kelly}.

We can test this value by integrating
the luminosity function and comparing the result to measurements
of the observed (by IRAS) total far-infrared emission of
nearby rich clusters.   \cite{hu} estimated
a 100$\mu$m emission of $310 \pm  110$ mJy for six clusters at a similar
average redshift of $z = 0.05$ but through a $10'$ diameter aperture.
This measurement corresponds to a 60$\mu$m flux density of $174 \pm 62$ mJy
(assuming the far infrared colors typical of a normal galaxy).
In the study of \cite{wise} the 27 clusters of richness class 1 or greater
(with an average of $z \sim  0.05$)
have an average $60\mu$m flux density of $201 \pm 157$ mJy in a $10'$ 
diameter beam.
To compare directly with the luminosity function
based estimates from the \cite{kelly} work,
we would need measurements in a larger beam, $\sim$ $30'$.
However, it is impossible to obtain IRAS
measurements with larger beams because noise
caused by confusion with the infrared cirrus 
increases with diameter more rapidly than does the 
signal from the galaxy cluster.
Nonetheless, it is difficult to reconcile the $\sim 2250$ mJy 
predicted in a 30' aperture by \cite{kelly} assuming a low end slope
of -0.8 with the $\sim 200$ mJy measured by \cite{hu} and \cite{wise}
in a 10' aperture.
For example, if such a steep aperture dependence held,
the clusters would be more readily detectable in a $30'$ beam by \cite{wise}.
In addition, \cite{kim_} show the Coma cluster is centrally 
concentrated in the radio source counts.
Finally, \cite{kelly} find the 60$\mu$m flux density of rich clusters at $z \sim 0.4$
to be approximately point-like in the $\sim$ $2'$ effective IRAS addscan beam.
Therefore, the data of \cite{hu} and \cite{wise} appear to be
in better agreement with the corrected flux densities of $\sim$ 800 mJy for
low end slopes of $-0.1$ or $-0.25$ than with that of $-0.8$ in \cite{kelly}.
In the following, we adopt a slope in this former range.

\cite{wang_} identify 5 galaxies detected with IRAS
within 3 Mpc (2 degrees) of the Coma cluster center
with $L_{IR} > 10^{10} L_\odot$.  Normalizing the cluster luminosity 
function to these five galaxies and using Eqns.~4 and 5
from \cite{kelly} to integrate it, we estimate that the total cluster luminosity
is $L_{IR} \sim 2 \times 10^{11} L_\odot$.

Since we expect a correlation between the H$\alpha$ emission
and far infrared luminosity we can compare the above total
luminosity to that predicted from the cluster galaxies
with measured H$\alpha$ luminosities.
\cite{gavazzi} measured
H$\alpha$ luminosities for a sample
of late-type galaxies with photographic magnitudes brighter than $m_p=15.4$.
The 13 late-type galaxies
within a degree of the cluster center observed by them have
H$\alpha$ luminosities of
$\sim 5\times 10^{40}$ erg/s corresponding to star formation
rates of $\sim 0.5 M_\odot$/yr and far infrared luminosities of
$\sim 2 \times 10^9 L_\odot$.
Assuming this luminosity as a limiting far infrared
magnitude and a total of 13 late-type galaxies
we derive a total cluster luminosity of $\sim 1.0 \times 10^{11}  L_\odot$, 
similar to but somewhat lower than that
estimated from the IRAS identifications.
This sample should give only a lower limit to the total
cluster luminosity because
it only covers late type galaxies
with $m_p < 16$ and consists only of optically bright spirals and so
may be missing optically faint but star forming late-type galaxies (such as D45)
as well as earlier-type galaxies.
Some, but not all, of the IRAS detected galaxies of \cite{wang_} 
are detected with bright H$\alpha$ emission by \cite{gavazzi}, 
emphasizing that the H$\alpha$ data give a lower limit to the infrared luminosity. 

The deep VLA study of \cite{kim_} is complete to a level
of 1.2mJy at 1.4GHz and covers a square degree centered on the
center of the Coma cluster. They estimate that there are $36\pm 5$  galaxies
with flux densities above 3mJy in excess of the expected background radio source counts.
At this level the radio source counts should be dominated by S0 and later
type galaxies undergoing star formation rather than
radio elliptical galaxies (e.g.,~\cite{ledlow}).
Assuming a typical far infrared to radio flux density ratio of 1/120
(\cite{hel85_}) the flux density limit of 3mJy at 1.4GHz corresponds
to a limiting luminosity of $3.4 \times 10^9 L_\odot$.
If we assume that 36 galaxies are in the cluster with luminosities
above this level, then the total luminosity of the cluster would be
$\sim 4.5 \times 10^{11} L_\odot$.
This value should be taken as an upper limit for a number of reasons.
For example, 50\% of the possible radio detections
are not optically identified and so have $m_p>19$;
their nature requires further investigation. A fraction of the
radio luminous galaxies are radio ellipticals and so should
not have been included in this tally because their radio emission 
may be dominated by sources not associated with recent star formation. 
Also the derived total cluster luminosity may be an upper limit
if the average FIR/radio ratio is lower than normal for galaxies
in dense clusters (\cite{anderson}).

In extrapolating to low luminosities, we have assumed that
there is no unique infrared emitting population that would
cause the luminosity function to differ from that in the field.
To test this assumption, we compare our estimates for the total
cluster luminosity to that which could be contributed from early type galaxies
such as the E+A ones studied here.
Our target galaxies were drawn from the \cite{cald93_} and \cite{cald97_}
survey of 182 early-type Coma cluster galaxies
out of which 18 had Balmer absorption, 3 had AGN-like characteristics
and 2 had emission lines typical of recent star formation.
20 galaxies total had abnormal spectral features since some had both
types of features. The sample was drawn from the complete (to B=20) sample of
\cite{godwin} of $\sim 800$ galaxies and was color and 
morphology selected to include all early-type (E and S0) galaxies.

Our sample of early-type galaxies with weak on-going star formation
has a limiting flux density $\sim 1$ mJy at 12$\mu$m.
Using a typical ratio of $F_{12\mu {\rm m}}/F_{60\mu {\rm m}} = 0.06$
this flux density corresponds to a limiting luminosity at $60\mu$m of
$L_{60 \mu {\rm m}} = 1 \times 10^8 L_\odot$.
We can assume the galaxies (6/8) detected at 12$\mu$m with abnormal
spectral features would have far infrared flux densities greater than this limit.
We then estimate that 15 galaxies would been found to have flux densities greater than 
this limit if we could have surveyed the entire list of
20 galaxies with abnormal spectral features. If 
we assume that this 20 represents all the relatively luminous early-type 
galaxies in the Coma cluster with abnormal spectral features, then there would be 15 galaxies total
with $L_{IR} \gtrsim 1 \times 10^8 L_\odot$.
Using Eqns 4 and 5 of \cite{kelly} but
with a luminosity function with a slope
at the faint end ranging from $-0.1$ to $-0.25$
we derive a total luminosity of $L_{IR} = 2 - 3 \times 10^{10} L_\odot$ for 
these galaxies, only a fraction of that predicted above for the full cluster. 
We conclude that early-type galaxies with weak ongoing
star formation are probably not major contributors to the
total far infrared cluster luminosity.

Typical nearby rich clusters contain less than 4\% of blue
late-type galaxies (\cite{rakos}). However,
the fraction of late-type galaxies in Coma
is estimated to be 14\% (\cite{dre80_}). The SW field contains
most of the galaxies with abnormal spectral features and is coincident
with a peak in X-ray emission  (and radio source counts; \cite{kim3_})
suggesting that a merger of subclusters has taken place
(\cite{cald93}). Since galaxies associated with this merger
probably make a significant
contribution to the total cluster far infrared flux density, the Coma cluster
luminosity should be higher than that of a typical rich nearby cluster.

\subsection{Origin of the far infrared emission of rich clusters}

\cite{dwek} modeled the distribution of dust that might be responsible for the reported 
extinction through the Coma cluster and found that the cluster would emit at 
$100\mu$m a flux density
of 0.2 MJy/sr, due to heating of this intergalactic dust by the hot cluster gas.
\cite{stickel} have reported detection of a surface
brightness of 0.1 MJy/sr at 120$\mu$m, which they take
to confirm this prediction. 
The technique used by Stickel et al.\ is based on a small (3\%) 
change in the ratio of 120$\mu$m to 185$\mu$m flux densities. 
Their data are not extensive enough to assess the variations in this ratio 
from Galactic infrared cirrus alone, which constitute the ``noise'' 
in their technique. In fact, their scan at PA 36$^o$ appears to be 
too noisy for an independent detection of the cluster. 
Thus, the reality of their detection is difficult to 
assess on simple signal to noise considerations.    

Moreover, in this paper we have found that the cluster galaxies themselves are
likely to be a significant source of far infrared emission.
It is noteworthy that we find roughly consistent estimates of
the 60 $\mu$m cluster luminosity in galaxies through three independent 
estimates: from the H$\alpha$ emission, we find a lower limit 
of $1 \times  10^{11} L_\odot$; from the radio emission
of individual galaxies, we find an
upper limit of $4.5 \times 10^{11} L_\odot$; and finally, 
directly from the infrared measurements of cluster galaxies, 
we deduce a luminosity of $2 \times 10^{11} L_\odot$. 
Assuming a luminosity of $2 \times 10^{11} L_\odot$ in the
central square degree of the cluster, and that the far infrared
flux density ratios are as in NGC~4853, we estimate a surface brightness of
0.06 MJy/sr at 100$\mu$m, which is similar to the 
excess source measured by \cite{stickel}. 
Thus, if the far infrared spectra of the Coma cluster galaxies 
are slightly hotter in 120/185 $\mu$m color than is 
the Galactic infrared cirrus, these galaxies could also 
be the cause of the apparent detection by \cite{stickel}. 

We conclude that the surface brightness
due to heating of the intergalactic dust is not convincingly 
detected and may be 
substantially lower than predicted by the model of \cite{dwek}.

\cite{wise} detected five of 56 clusters in the
IRAS far infrared data. In two cases, the emission was possibly extended
and they concluded it might arise in the intracluster medium. 
In other cases, the detected flux density could
be attributed to individual galaxies.
\cite{hu} measured a modest level of extinction in cooling flow clusters
and argued that a far infrared emission of $\sim 0.5$ Jy at 100$\mu$m
was expected due to heating of the
intergalactic dust responsible for this extinction.
She found an average flux density of 0.3 Jy from the IRAS data,
in agreement with this prediction.

In both these cases, we have
shown that the far infrared emission of the clusters is
in reasonably good agreement with the flux densities that \cite{kelly}
found for similar clusters, if we update their result with an improved 
galaxy luminosity function.  Because the flux densitiess of \cite{kelly} were obtained
by summing the output of individual galaxies,
this agreement suggests, as we found for Coma, that much of the far infrared
emission of these clusters arises from their
member galaxies and not from the intergalactic dust.

\subsection{Luminosity evolution}

We now compare the typical 60$\mu$m cluster flux density
of 800 mJy at $z = 0.05$,
(corrected from \cite{kelly_}) to
the measurement of \cite{kelly} of an average flux density
of 28 mJy at $z = 0.4$ for the distant cluster sample.
At 60$\mu$m the K correction between these two redshifts is about
a factor of $0.6$ (\cite{kelly_}),
so the predicted 60$\mu$m flux density at $z = 0.4$ would be 7.5mJy
assuming no luminosity evolution.
By comparing the measurements at the two redshifts we find
a nominal luminosity evolution that is proportional to $(1+z)^{3.9}$.
The uncertainties in this estimate remain large because
of the problems in estimating the far infrared luminosity of
the nearby clusters.
\cite{saunders97} find a far infrared density evolution for field
galaxies in the IRAS data of $(1+z)^{(3.7 \pm 1.2)}$,
which is equivalent to a luminosity evolution of about $(1+z)^2$.
Thus, if the far infrared emission from rich clusters arises
primarily from their member galaxies, as we argue in the previous section,
then we infer that the luminosity evolution of these galaxies is similar to,
or perhaps somewhat stronger than, the evolution of field galaxies.

\section{Conclusions}

We have presented broad band 12$\mu$m images of 9 Coma cluster
galaxies, 8 of which show either E+A characteristics or/and
narrow emission lines. We have detected 7 galaxies,
5 with narrow emission lines (including 2 AGNs),
one bright control galaxy
lacking abnormal spectral characteristics, and one pure E+A galaxy
lacking narrow emission lines but detected at the low flux density 
of 0.4 mJy.
The remaining two E+A galaxies (lacking narrow emission lines)
were not detected with an upper limit of 0.3 mJy.
We use these data and information from the literature to conclude:

\noindent
1.) Only galaxies with emission lines have significantly enhanced
12$\mu$m flux densities compared to their 2.2 micron flux densities.
The E+A phase (lacking emission lines) appears to be truly a post starburst
phase with little on-going star formation.

\noindent
2.) The one galaxy in our sample also detected by IRAS, NGC~4853,
has similar mid and far-infrared colors similar
to those of a normal spiral galaxy.  This behavior lets us
estimate far infrared outputs for galaxies with
enhanced 12$\mu$m flux densities.

\noindent
3) By comparing the current star formation rates
with previous rates estimated from the Balmer absorption features
we divide the galaxies into two groups:
those having current star formation rates that
are substantially lower than that required to form their $\sim 1$ Gyr old
populations (NGC~4853, D94, D99, D44 and D112)
and those with significant ongoing star formation
or/and AGN activity (D45, D100, and D16).
We note no strong difference in the
spatial distribution on the sky between these two groups, however
they have distinct ranges for their systemic velocities.
The first group has velocities above the
mean cluster velocity and the second group has velocities
well below it.   This would be consistent with the interpretation
of \cite{colless_} that the different populations in the Coma
cluster originate from distinct cluster infall events.
The kinematic differences between the two groups suggests that
star formation triggered by the cluster decays on a timescale of order a Gyr.

\noindent
4.) Using the IRAS identifications in the Coma cluster (\cite{wang}),
the H$\alpha$ survey of late-type galaxies by \cite{gavazzi}  and the radio survey of \cite{kim_}, and summing the indicated the far infrared luminosity
function, we estimate that the total far infrared luminosity of the
Coma cluster is $\sim 2 \times 10^{11} L_\odot$.

\noindent
5.) The early-type galaxies with abnormal spectral
features (Balmer absorption or narrow line emission)
contribute only a small fraction of the total luminosity.
Late type galaxies, such as those that tend to be infrared bright in the field,
dominate the Coma cluster galaxian luminosity.
Their integrated luminosity is a substantial portion,
probably the majority, of the far infrared luminosity of the cluster.

\noindent
6.)  The far infrared emission of the Coma cluster intergalactic medium
is not yet reliably detected. A complication in measuring this emission in large beam observations that do not resolve the individual galaxies is the potentially dominant role of the galaxies in the integrated cluster far infrared emission. 

\noindent
7.)  Previous indications of very little far infrared cosmic
luminosity evolution in rich clusters probably are incorrect
because of an overestimate of the average luminosity of nearby clusters.
Using an updated galaxy luminosity function,
we find that the cluster evolution is as fast as,
or perhaps even faster than, that of field galaxies.
This result remains uncertain because of the difficulties
in determining integrated luminosities for nearby rich clusters.

\acknowledgments
We acknowledge helpful discussions and correspondence with
Frazer Owen, Wayne Barkhaus and Doug Kelly.
We thank D.~van Buren, M.~Seh, K.~Ganga, R.~Hurt,
L.~Hermans and the ISO team at
IPAC for help with the data reduction of the ISOCAM images.
We also acknowledge support from NSF grant AST-9529190
and NASA project no.~NAG-53359.

{}

\clearpage

\begin{figure*}
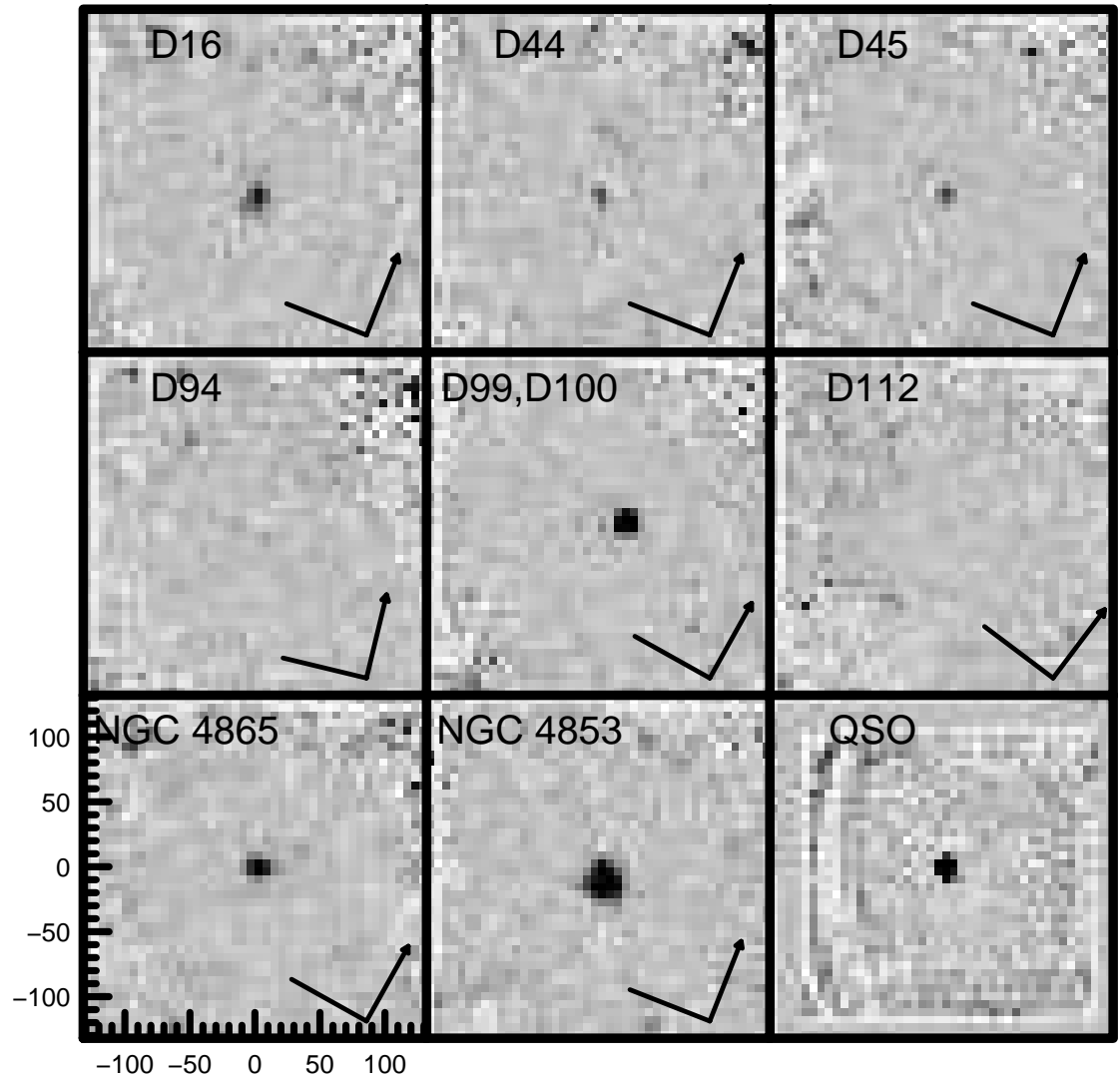

\caption[junk]{
12$\mu$m (LW10 filter) ISOCAM broad band images of Coma cluster galaxies.
In the image centered on D99 we are seeing emission from D100,
not D99. The arrow in the lower right corner of each image
points towards north and the other axis towards east.
The scale in arcsecs is given on the lower left image.
For comparison a point source
is shown in the lower right image.
This quasar has a flux density of 10mJy (similar to D100)
and was observed with a similar exposure time
and the same camera, filter and pixel size as the galaxies.
\label{fig:fig1} }
\end{figure*}

\clearpage


\begin{deluxetable}{lrrccccccccr}
\footnotesize
\scriptsize
\tablenum{1}
\tablecaption{Coma Cluster Galaxies}
\tablehead{
\multicolumn{1}{l}{Galaxy}   &
\multicolumn{1}{r}{Features}           &
\colhead{$F_{12\mu{\rm m}}$} &
\colhead{K}                  &
\colhead{J-K}                &
\colhead{$F_{2.2\mu{\rm m}}/F_{12 \mu{\rm m}}$}  &
\colhead{B}                  &
\colhead{B-R}                &
\colhead{$L_{IR}/ L_{opt}$}  &
\colhead{$L_{IR}$}           &
\colhead{$V_{helio}$}         \nl
\colhead{}     &
\colhead{}     &
\colhead{mJy}  &
\colhead{mag}  &
\colhead{mag}  &
\colhead{}     &
\colhead{mag}  &
\colhead{mag}  &
\colhead{}     &
\colhead{$10^8 L_\odot$}  &
\colhead{km/s}   \nl
\multicolumn{1}{l}{(1)}  &
\multicolumn{1}{r}{(2)}  &
\colhead{(3)}  &
\colhead{(4)}  &
\colhead{(5)}  &
\colhead{(6)}  &
\colhead{(7)}  &
\colhead{(8)}  &
\colhead{(9)}  &
\colhead{(10)} &
\colhead{(11)}
}
\startdata
NGC 4853&N A&$39.8\pm  1.2$ &10.84 &1.01 & 0.7  &14.38 &1.66 & 0.59& 74& 7660\nl
D100    &N A&$9.31\pm 0.62$ &12.84 &0.92 & 0.5  &16.25 &1.41 & 0.77& 11& 6513\nl
D16  &N  ~ &$4.53\pm 0.48$ &12.22 &0.97 & 1.8  &16.14 &1.82 & 0.34& 8.3&6205\nl
D45     &N A&$1.69\pm 0.38$ &14.61 &0.72 & 0.5  &16.72 &1.80 & 0.22& 3.1&4915\nl
D44     &N A&$1.21\pm 0.40$ &13.05 &0.84 & 3.1  &16.57 &1.77 & 0.14& 2.2&7534\nl
D112    &  A&$0.4 \pm 0.25$ &13.02 &0.93 & 10   &16.64 &1.78 &     &   &7428 \nl
D94     &  A&$<0.3        $ &13.21 &0.83 &$>10 $&16.49 &1.62 &     &   &7084 \nl
D99     &  A&$<0.3        $ &13.18 &0.85 &$>10 $&16.98 &1.53 &     &   &9902 \nl
NGC 4865&   &$5.95\pm 0.54$ &10.65 &0.97 & 5.8  &14.54 &1.83 &     &   &4609 \nl
\enddata
\tablenotetext{}{NOTES.-- Columns:
(1) When no NGC number exists,
galaxies are referred to by their Dressler (1980) or D number;
(2) A denotes the presence of Balmer absorption lines
and N denotes the presence of narrow emission lines.
We note that D16 had a high exitation spectrum and a broad H$\alpha$ component
therefore is probably
a Seyfert 1 galaxy (Caldwell et al.~1996).
D44 also has a spectrum with exitation between a LINER and
a Seyfert 2 (Caldwell et al.~1996).  D45, D100 and NGC~4853
have emission line spectra typical of HII regions;
(3) 12$\mu$m flux densities in mJy measured from the LW10 filter in a $60''$
diameter aperture.
A photometric error is estimated from the signal to noise. The absolute
calibration may also be uncertain by $\pm 20\%$;
(4) The K magnitude is measured in a $36''$ diameter aperture from the
Ks images with nearby stars and galaxies removed.
The absolute estimated error is $\sim 0.1$ magnitude, though the relative
error between galaxy measurements is smaller ($\sim 0.05$ magnitude);
(5) The J-K color is measured in a $9.6''$ diameter aperture;
(6) 2.2$\mu$m to 12$\mu$m flux density ratio;
(7,8) B magnitudes and B-R colors are photographic and are
taken from Godwin et al.~(1983);
(9,10) $L_{IR}/ L_{opt}$ is the infrared to optical
luminosity ratio and $L_{IR}$ is the estimated far infrared luminosity
in units of $10^8 L_\odot$.
$L_{opt}$ is computed as $\nu F_\nu$ at B band
and $L_{IR}$ similarly but at 60$\mu$m from the 12$\mu$m flux density using
the flux density ratio $F_{12 \mu{\rm m}} / F_{60 \mu {\rm m}} = 0.06$.
We have not estimated $L_{IR}$ for the E+A galaxies D94, D99
and D112 or the normal galaxy NGC~4865 since they have
2.2 to 12$\mu$m flux density ratios
typical of an old stellar population (or early-type galaxy),
and so we have no direct estimate for the 12 to 60$\mu$m flux density ratio;
(11) Systemic velocities are taken from Caldwell et al.~(1993).
}
\end{deluxetable}


\clearpage
\begin{thebibliography}{}


\bibitem[Anderson \& Owen 1995]{anderson}
Anderson, V., \& Owen, F.~N.~1995, AJ, 109, 1582

\bibitem[Briel, Henry \& Bohringer 1992]{briel}
Briel, U.~G., Henry, J.~P., \& Bohringer, H.~1992, A\&A, 259, L31

\bibitem[Butcher \& Oemler 1978]{but78}
Butcher, H., \& Oemler, A.\ 1978, ApJ, 219, 18

\bibitem[Butcher \& Oemler 1984]{but84}
Butcher, H., \& Oemler, A.\ 1984, ApJ, 285, 426

\bibitem[Caldwell \& Rose 1997]{cald97}
Caldwell, N., \& Rose, J.~A.\ 1997, AJ, 113, 492

\bibitem[Caldwell et al.~1996]{cald96}
Caldwell, N., Rose, J.~A, Franx, M., \& Leonardi, A.~J.\ 1996, AJ, 111, 78

\bibitem[Caldwell et al.~1993]{cald93}
Caldwell, N., Rose, J.~A, Sharples, R.~M., \& Bower, R.~G.\ 1993, AJ, 106, 473

\bibitem[C\'esarsky et al.~1996]{ces96}
C\'esarsky, C.~J.\ et al.\ 1996,  A\&A, 315, L32


\bibitem[Colless \& Dunn 1996]{colless}
Colless, M., \& Dunn, A.~M.~1996, ApJ, 458, 435

\bibitem[Couch \& Sharples 1987]{cou87}
Couch, W., \&  Sharples, R.~M.\  1987, \mnras,  229, 423

%
\bibitem[Couch et al.~1998]{cou98}
Couch, W.~J., Barger, A.~J., Smail, I., Ellis, R.~S., \& Sharples,
R.~M.\ 1998, ApJ, 497, 188

\bibitem[DeJong et al.~1984]{dejong}
DeJong, T., Clegg, P.~E., Soifer, B.~T., Rowan-Robinson, M., Habing, H.~J.,
Houck, J.~R., Aumann, H.~H., \& Raimond, E.~1984, ApJ, 278, L67

\bibitem[Devereux \& Young 1990]{dev90}
Devereux, N.~S., \& Young, J.~S.~1990, ApJ, L25

\bibitem[Dressler (1980)]{dre80}
Dressler, A.~1980, ApJS, 42 565

\bibitem[Dressler 1984]{dre84}
Dressler, A.~1984, ARA\&A, 22, 185

\bibitem[Dressler \& Gunn 1982]{dre82}
Dressler, A., \& Gunn, J.~E.\ 1982, ApJ, 263, 533

\bibitem[Dressler \& Gunn 1983]{dre83}
Dressler, A., \& Gunn, J.~E.\ 1983, ApJ, 270, 7

\bibitem[Dressler et al.~1994]{dre94}
Dressler, A., Oemler, A., Butcher, H.~R., \& Gunn, J.~E.~1994,
ApJ, 430, 107

\bibitem[Dwek et al.~(1990)]{dwek}
Dwek, E., Rephaeli, Y., \& Mathur, J.~1990, ApJ, 475, 565

\bibitem[Elias et al.~(1982)]{elias}
Elias, J.~H., Frogel, J.~A., Matthews, K., \& Neugebauer, G.\ 1982,
\aj, 87, 1031

\bibitem[Gavazzi et al.~(1998)]{gavazzi}
Gavazzi, G., Catinella, B., Carrasco, L., Boselli, A., \&
Contursi, A.~1998, AJ, 115, 1745

\bibitem[Godwin, Metcalfe \& Peach (1983)]{godwin}
Godwin, J.~G., Metcalfe, N., \& Peach, J.~V.\ 1983, \mnras, 202, 113

\bibitem[Helou 1986]{hel86}
Helou, G.\ 1986, \apj, 311, L33

\bibitem[Helou et al.~(1988)]{hel88}
Helou, G., Khan, I.~R., Malek, L., \& Boehmen, L.\ 1988, \apjs, 68, 151

\bibitem[Helou et al.~(1985)]{hel85}
Helou, G., Soifer, B.~T., \& Rowan-Robinson, M.~1985, ApJ, 298, 7

\bibitem[Hu (1992)]{hu}
Hu, E.~M.~1992, ApJ, 391, 608

\bibitem[Impey et al.~1986]{imp86}
Impey, C.~D., Wynn-Williams, C.~G., \& Becklin, E.~E. 1986, \apj, 309, 572

\bibitem[Kelly \& Rieke (1990)]{kelly}
Kelly, D.~M., \& Rieke, G.~H.\ 1990, ApJ, 361, 354

\bibitem[Kennicutt 1983]{kenn83}
Kennicutt, R.~C., Jr.~1983, ApJ, 272, 56

\bibitem[Kessler et al.~1996]{kes96}
Kessler, M.~F., Steinz, J.~A., Anderegg, M.~E., Clavel, J.,
Drechsel, G., Estaria, P., Faelker, J., Riedinger, J.~R.,
Robson, A., Taylor, B.~G., \& Xime\'nez de Ferra\'n, S.\ 1996, A\&A, 315, L27

\bibitem[Kim et al.~1994a]{kim}
Kim, K.~-T., Kronberg, P.~P., Dewdney, P.~E., \& Landecker, T.~L.~1994a,
A\&AS, 105, 385

\bibitem[Kim et al.~(1994b)]{kim3}
Kim, K.~-T., Kronberg, P.~P., Dewdney, P.~E., \& Landecker, T.~L.~1994b,
A\&A, 288, 122

\bibitem[Knapp et al.~1992]{kna92}
Knapp, G.~R., Gunn, J.~E., \& Wynn-Williams, C.~G.
1992, \apj, 399, 76

\bibitem[Ledlow \& Owen 1996]{ledlow}
Ledlow, M.~J., \& Owen, F.~N.~1996, AJ, 112, 9

\bibitem[Leitherer \& Heckman 1995]{leth95}
Leitherer, C., \& Heckman, T.~M.\ 1995, ApJS, 96, 9

%


\bibitem[Moore et al.~1996]{moore96}
Moore, B., Katz, N., Lake, G., Dressler, A., \& Oemler, A.~1996,
Nature, 379, 613

\bibitem[Moore, Lake \& Katz 1998]{moore}
Moore, B., Lake, G., \& Katz, N.~1998, ApJ, 495, 139


\bibitem[Oemler, Dressler \& Butcher 1997]{oem97}
Oemler, A., Jr., Dressler, A., \& Butcher, H.~R.~1997
ApJ, 474, 561

\bibitem[Rakos \& Schombert 1995]{rakos}
Rakos, K.~D., \& Schombert, J.~M.~1995, ApJ, 439, 47

\bibitem[Roberts et al.~1991]{rob91}
Roberts, M.~S., Hogg, D.~E., Bregman, J.~N., Forman, W.~R., \&
Jones, C.\ 1991, ApJS, 75, 751

\bibitem[Saunders et al.~1990]{saunders}
Saunders, W.,  Rowan-Robinson, M., Lawrence, A., Efstathiou, G.,
Kaiser, N., Ellis, R.~S.,\&,  Frenk, C.~S. 1990, \mnras, 242, 318

\bibitem[Saunders et al.~(1997)]{saunders97}
Saunders, W.~et al.~1997, in
``Extragalactic Astronomy in the Infrared,'' ed. Mamon,
Thuan, \& Van, Editions Frontiers, p431

\bibitem[Soifer et al.~1986]{soi86}
Soifer, B.~T., Rice, W.~L., Mould, J.~R., Rowan Robinson, M., \& Habing, H.~J.
1986, \apj, 304, 351

\bibitem[Soifer et al.~1989]{soi89}
Soifer, B.~T., Boehmer, L., Neugebauer, G., \& Sanders, D.~B.\ 1989,
\aj, 98, 766

\bibitem[Sparke, Kormendy \& Spinrad (1980)]{sparke}
Sparke, L.~S., Kormendy, J., \& Spinrad, H.\ 1980, ApJ, 235, 755

\bibitem[Spinoglio et al.~(1995)]{spinoglio}
Spinoglio, L., Malkan, M.~A., Rush, B., Carrasco, L., \& Recillas-Cruz,
E.~1995, 453, 16

\bibitem[Stickel et al.~(1998)]{stickel}
Stickel, M., Lemke, D., Mattila, K., Haikala, L.~K., \& Hass, M.\ 1998,
A\&A, 329, 55

\bibitem[Thompson \& Gregory (1993)]{thompson}
Thompson, L.~A., \& Gregory, S.~A.~1993, AJ, 106, 2197

\bibitem[Vigroux, Boulade \& Rose 1989]{vigroux}
Vigroux, L., Boulade, O., \& Rose, J.~A.~1989, AJ, 98, 2044

\bibitem[Wise et al.~(1993)]{wise}
Wise, M.~W., O'Connell, R.~W., Bregman, J.~N., \& Roberts, M.~S.~1993,
ApJ, 405, 94

\bibitem[Wang et al.~1991]{wang}
Wang, G., Clowes, R.~G., Leggett, S.~K., MacGillivray, H.~T., \&
Savage, A.~1991, \mnras, 248, 112

\bibitem[Worthey 1993]{worthey}
Worthey, G. 1994, ApJS, 95, 107

\bibitem[Zabludoff et al.~1996]{zabludoff}
Zabludoff, A.~I., Zaritsky, D., Lin, H., Tucker, D., Hashimoto, Y.,
Shectman, S.~A., Oemler, A., \& Kirshner, R.~P.~1996,
\apj, 466, 104

\bibitem[Kelly \& Rieke 1990]{kelly_}
\bibitem[Caldwell et al.~(1993)]{cald93_}
\bibitem[Caldwell et al.~(1996)]{cald96_}
\bibitem[Caldwell \& Rose (1997)]{cald97_}
\bibitem[Devereux \& Young (1990)]{dev90_}
\bibitem[Wang et al.~(1991)]{wang_}
\bibitem[Kim et al.~1994b]{kim3_}
\bibitem[Kim et al.~(1994a)]{kim_}
\bibitem[Spinoglio et al.~1995]{spinoglio_}
\bibitem[Helou et al.~1985]{hel85_}
\bibitem[Dressler 1980]{dre80_}
\bibitem[Stickel et al.~1998]{stickel_}
\bibitem[Wise et al.~1993]{wise_}
\bibitem[Hu 1992]{hu_}
\bibitem[Gavazzi et al.~1998]{gavazzi_}
\bibitem[Colless \& Dunn (1996)]{colless_}

\end{thebibliography}
\end{document}